\documentstyle[preprint,aps,eqsecnum]{revtex}
\begin{document}
\draft
\title{
Heavy diquark in baryons containing a single heavy quark and the weak form
factors}
\author{
Le Viet Dung \cite{LVD}}
\address{
 International Centre for Theoretical Physics, Trieste 34100, Italy}
\author{ Nguyen Ai Viet}
\address{ Physics Department, Syracuse University, Syracuse, New York 13244}
\maketitle
\begin{abstract}
It is shown that the number of independent weak form factors collapses if
the heavy diquark exists inside the baryons containing a single heavy quark.
The relations between the weak form factors are quite different in the case
of light diquark.
So a careful analysis on the future data of the weak form factors would clarify
that in those baryons the correlation between
the heavy quark and a light quark is stronger or weaker than the one between
two light quarks.
\end{abstract}
\newpage

\section{Introduction}
In the limit the heavy quark mass going to infinity, a new $SU(2N_f)$ ($N_f$
is number of heavy flavors ) spin-flavor symmetry appears in QCD\cite{HQET}.
As a
consequence, this symmetry implies that all the form factors involved in the
semi-leptonic decays of $B$ meson can be expressed by one single universal
Isgur-Wise function.
Recently\cite{HQ}, it has been argued that in the baryons containing
a single heavy quark ( hereafter we shall refer to as 1HQ-baryons) the
correlation between the heavy quark and one light quark is much stronger than
the one between two light quarks. That is to say, the presence of a heavy
diquark
could be more probable than the one of a light diquark in 1HQ-baryons.
The main arguments are follows: The gluons that the heavy quark exchanges with
light degrees of freedom are soft only for the heavy quark but are extremely
hard for the light degrees of freedom ( their momenta are of order
$ \sqrt {M_Q.\Lambda _{QCD} } $\cite{VOSHI}). Suppose that at some moment
there exists
some light diquark. The dynamics also contains hard gluon exchanges which have
enough energy to break the binding of the light diquark, which is of order
$\Lambda _{QCD}$. If we integrate the hard
gluons into the diquark to have a strong binding of the diquark and a weak
interaction of it with the heavy quark, we must have a very large diquark
which would loose the sense of its own use. On the other hand, the heavy
diquark
must be very stable, since the hardest gluons have been integrated into the
diquark to build a strong binding. The remaining light diquark has not enough
energy to break this binding. The similar approximation has been applied in
atomic physics: The $He$ atom can be well approximated as a system of an
electron with a $He^+$ ion. It could hardly be considered as a $He^{++}$ ion
and a Cooper pair of two electrons. Some calculations\cite{COR} have also
indicated that
in the 1HQ-baryons the correlation of qq is always much weaker than the one of
$Qq$ ($ q $ denotes light quarks, while $Q$ denotes heavy quarks).

So far, we are no fanatics of the idea and of the heuristic arguments given
above,
unless it could be proved from the first principles of QCD. Such
nonperturbative
aspects of QCD are still exceedingly difficult to attack at present. We need
more evidences to headaway the idea, which would lead to an interesting
supersymmetry between 1HQ-mesons and baryons[2]. Fortunately, in this paper we
are
able to offer a method to analyze the weak decays of 1HQ-baryons, which may
likely trace back to the presence of the existing diquark. We shall show that
the diquark, if exists, would leave its traces on a series of relations between
the weak form factors of baryons. Quite different relations are implied from
the
presumed existence of the heavy diquark or of the light diquark. So a careful
analysis of the baryon's weak decay data in the future will clarify the
dominant correlation in 1HQ-baryons.

In such an analysis a proper definition of diquark is very essential. Although
the idea of diquark was suggested as early as in 1966\cite{DIQ}, there are
still
many contradictory definitions of diquark\cite{LICH}. Many papers contain the
assumption, sometimes hidden, that diquarks can be treated as elementary. Some
others state that diquarks are quasi-elementary constituents of baryons, or
even
assume that diquarks are not elementary at all but rather correlated states of
two quarks. In this paper we shall use
another definition of diquark, which would satisfy the requirement of
Lichtenberg\cite{LICH}: a diquark is a correlated state of two quarks and the
quarks
inside and outside the diquark should be antisymmetrized. To give the diquark
a sense we shall also assume that the diquark spin is a well-defined quantum
number. This is the case when the spin-spin interaction between the diquark and
the remaining degrees of freedom is small comparing to the color force. In such
a strong regime of QCD, the diquark spin is a good quantum number (while the
spin
of different quark can be not a good quantum number ). Therefore, we
can split the spin part of the baryon wave functions respectively. In the large
limit of the heavy quark mass, we can also postulate the heavy quark symmetries
for 1HQ-baryons. So, the heavy diquark picture can be considered as a further
step
beyond the spectator quark model. Such an approximation is similar to the one
when we approximate an atom by an electron bound to an ion neglecting the
electron-electron interaction. We shall follow the covariant Bethe-Salpeter
formalism \cite{HU1}
and show that the assumption of heavy diquark reduces the number of the weak
form factors of 1HQ-baryons  comparing to the cases where the light diquark or
the heavy quark symmetry alone are supposed.

The paper is organized as follows: In Sec.II we shall briefly review
the covariant Bethe-Salpeter formalism and the weak current induced baryon
transitions. In Sec.III giving a proper definition of diquark we
develop the covariant Bethe-Salpeter formalism for the heavy diquark-quark
systems to construct the wave functions of heavy baryons. The heavy to heavy
and the heavy to light baryon transitions are discussed in Secs.IV and
V, respectively. In Sec.VI we shall discuss the results and suggest
some new possibilities opened by those ones. The light diquark-quark picture of
1HQ-baryons does not lead to any simplification since actually the spin
decoupling of the heavy quark coincides to the one of the light diquark.

\section{The covariant Bethe-Salpeter formalism and the weak current induced
transitions of the 1HQ-baryons}
Within the covariant constituent quark model, the baryon Bethe-Salpeter
amplitude is:
\begin{equation}
 B = < 0 | T(\psi_\alpha (x_1)\psi_\beta (x_2)\psi_\gamma(x_3) | B >
\end{equation}
where the $\psi$'s represent the quark fields and $ | B >$ is a baryon state.
The B-S amplitude contains the baryon-quark vertex function with quark and
baryon legs. The Feynman rules are given by quark propagators and the truncated
B-S amplitude\cite{HU2}.

 Diagramatically, the B-S amplitude is represented in Fig.1.\\
\vskip 3truecm
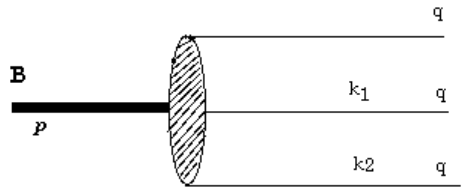
\centerline { {\footnotesize Fig.1}}

The Bethe-Salpeter wavefuntions of baryons should be fully antisymmetrized
under
the interchange of any two quarks. So the Fourier transform of the fully
antisymmetric baryon B-S wave function is represented as follows:
\begin{equation}
B_{A_i;B_j;C_k} =  { \epsilon_{ijk} \over \sqrt {6}} B_{(A;B;C)}(P;k_1,k_2)
\end{equation}
where $ A=( \alpha, a )$, etc. represents the Dirac and flavor indices
$\alpha $ and $ a$, respectively. The color labels are denoted by i, j, k. The
variables $k$ and $P$ denote the quark and baryon momenta respectively. By the
conservation law we can allow the baryon wave function having only three
arguments $k_1,k_2 $ and $P$. The spin-flavor part $ B_{(ABC)}$ should be fully
symmetric under the interchange of the indices $ A,B,C$.
 By help of the spin-parity projectors
we can project out the irreducible states\cite{FEDU,HU1} of spin $s={3\over
2}$ and $s={1\over 2}$.

The natural ansatz for the baryon Bethe-Salpeter wave functions in general is
given as follows:
\begin{equation}
 B_{(A;B;C)}(P;k_1,k_2) = \chi^B_{\rho \delta \sigma ; abc} A^{\rho \delta
\sigma}_{\alpha \beta \gamma} (P;k_1,k_2)
\end{equation}
Here $A^{\rho \delta \sigma }_{\alpha \beta \gamma }(P;k_1,k_2)$ is a
sixth-rank
Dirac tensor function, while $\chi^B_{\rho \delta \sigma ;abc}$ is the
spin-parity
flavor projector.

At the moment we ignore the flavor-dependent part having in mind that it is
determined uniquely by the spin-dependent part and it can be constructed
conveniently, whenever it will be necessary. In general, the spin-parity
projectors for the s-wave baryons are given as follows\cite{HU1}:
\begin{equation}
 \chi^\Lambda _{\rho \delta \sigma } =  u_\rho [(\not v +1)\gamma _5 C]_{\delta
\sigma }
\end{equation}
\begin{equation}
 \chi^\Sigma_{\rho \delta \sigma } = -[((\gamma^\mu +v^\mu)\gamma _5 u]_\rho
[(\not v + 1)\gamma _\mu C]_{\delta \sigma }
\end{equation}
\begin{equation}
 \chi^{\Sigma ^*}_{\rho \delta \sigma } = u^\mu _\rho [(\not v +1)\gamma _
\mu C]_{\delta \sigma }
\end{equation}
where $v_\mu = P_\mu /M_B $ is the four-velocity of baryons. The Dirac spinors
$u_\alpha $ satisfy the equation:
\begin{equation}
(\not v -1) u = 0
\end{equation}
The Rarita-Schwinger spinors $u^\mu_\alpha $ satisfy the equations:
\begin{equation}
(\not v -1) u^\mu
\end{equation}
\begin{equation}
\gamma _\mu u^\mu =0
\end{equation}
\begin{equation}
 v_\mu u^\mu = 0
\end{equation}
The general wave functions given in the formula (2.3) can be used to describe
the light baryons. For the 1HQ-baryons,
it has been shown that if one of quarks becomes heavy, in the limit of the
heavy quark mass going to infinity, we have the following good approximation:
\begin{equation}
 A^{\rho \delta \sigma }_{\alpha \beta \gamma } \approx \delta^\rho_\alpha
A^{\delta\sigma}_{\beta\gamma}(v,k_1,k_2)
\end{equation}
Diagramatically, we can represent the 1HQ-baryon- quark Bethe-Salpeter
amplitude
as in Fig.2.\\
\vskip 3truecm
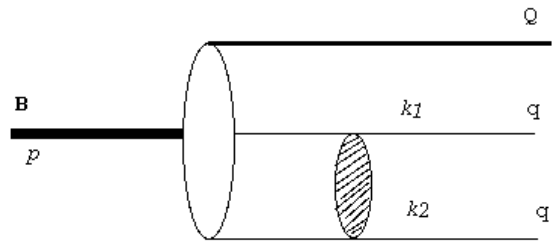
\centerline{ {\footnotesize Fig.2 }}
Consider the weak current induced baryon transitions, which are represented
diagramatically in Fig.3.
\vskip 3truecm
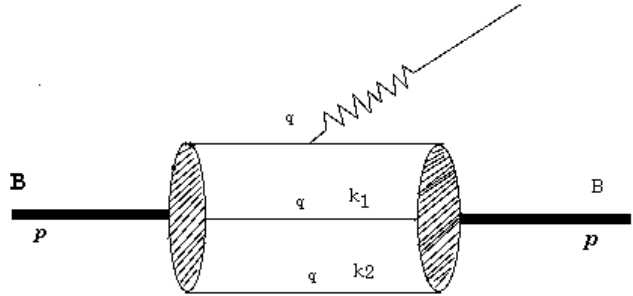
\centerline {{\footnotesize Fig.3 }}
The Feynman rules are given by the quark propagators and the truncated
quark-baryon Bethe-Salpeter amplitudes as vertex functions. Therefore, the
decay
matrix element can be written as
\begin{equation}
 < B_2(P_2) | J_\lambda^{V-A} | B_1(P_1) > = \int d^4k_1 d^4k_2 \bar B^{\alpha
\beta\gamma} [\gamma_\lambda(1-\gamma_5)]_\alpha^{\alpha'}(\not k_1 - m_1)^
{\beta'}_\beta
\end{equation}

\begin{equation}
 (\not k_2 -m_2)^{\gamma'}_\gamma B_{\alpha' \beta'\gamma'}.b(B_1,B_2)
\end{equation}
We have already absorbed the Kobayashi-Maskawa-Cabbibo matrix element into the
trace of the flavor part $ b(B_1,B_2)$. Having an ansatz of
the wave functions we can compute the decay matrix for different transitions.
The results are follows (see Refs.\cite{HU2} for more details).
\subsection {{\it The heavy to heavy baryon transitions}}
Using the baryon Bethe-Salpeter wave function projected out by the spin-parity
projectors (2.4-6) and the
approximation (2.11) we can write down the decay matrix elements as follows:\\

i) $\Lambda (\Xi )$-type ${1\over 2}^+ \rightarrow {1\over 2}^+$ baryon
transitions
\begin{equation}
< \Lambda_2(P_2) | J^{V-A}_\lambda | \Lambda_1(P_1) > = b(\Lambda_1,\Lambda_2)
\bar u_2(P_2) \gamma_\lambda (1-\gamma_5) u_1(P_1).F_\Lambda
\end{equation}

ii) $\Sigma (\Omega)$-type ${1\over 2}^+ \rightarrow {1\over 2}^+$ transitions
\begin{eqnarray}
<\Sigma_2(P_2) | J_\lambda^{V-A} | \Sigma_1(P_1) >& = & b(\Sigma_1,\Sigma_2)
\bar u_2(P_2) \gamma_5 (\gamma^\mu + v_2^\mu) \gamma_\lambda (1-\gamma_5)
(\gamma^\nu + v^\nu_1)\nonumber\\
\gamma_5 u_1(P_1) (F_1 g_{\mu \nu } & + & F_2 v_{1\mu } v_{2\nu })=
 2 b(\Sigma_1,\Sigma_2)\bar u_2(P_2)\Big[ F_L\gamma_\lambda (1- \gamma_5
)\nonumber \\
-{2\over 1+\omega } (F_L + F_T )( v_{1\lambda } + v_{2\lambda }) & + &
{2\over 1-\omega } ( F_L - F_T ) ( v_{2 \lambda
} - v_{1\lambda })\gamma_5\Big] u_1(P_1)
\end{eqnarray}

iii) $\Sigma (\Omega ) \rightarrow \Sigma^*(\Omega^*)$-type ${1\over 2}^+
\rightarrow {3\over 2}^+$ transitions
\begin{eqnarray}
<\Sigma_2^*(P_2) |& J_\lambda^{V-A}& | \Sigma_1(P_1) > = b(\Sigma_1,\Sigma^*_2)
\bar u^\mu_2 \gamma_\lambda ( 1- \gamma_5) (\gamma^\nu + v^\nu_1) \gamma_5 u_1
(P_1)\Big ( F_1 g_{\mu \nu }
+ F_2 v_{1\mu } v_{2\nu }\Big) \nonumber\\
&=& b(\Sigma_1,\Sigma_2^*)\bar u_2^\mu (P_2)\Big[ F_T g_{\mu \lambda }
( 1+ \gamma_5) + {1\over 2(1+\omega )} (F_L
 + F_T ) v_{1\mu }\gamma_\lambda
\gamma_5 \nonumber\\
& - & {1\over 1-\omega^2}(F_L - \omega F_T ) v_{1\mu } v_{2\lambda }
(1+\gamma_5)\Big]u_1(P_1)
\end{eqnarray}

iv) $\Sigma^*(\Omega^*)\rightarrow \Sigma^*(\Omega^*)$-type ${3\over 2}^+
\rightarrow {3\over 2}^+$ transitions
\begin{eqnarray}
< \Sigma_2^*(P_2) | &J^{V-A}_\lambda & | \Sigma_2^*(P_1) >\nonumber \\
& = & b(\Sigma^*_1,\Sigma_2^*) \bar u_2^\mu (P_2) \gamma_\lambda (1- \gamma_5)
u^\nu_1(P_1) (F_1 g_{\mu \nu } + F_2 v_{1\mu } v_{2\nu })\nonumber \\
& = & 4 b(\Sigma_1^*,\Sigma_2^*) \bar u_2^\mu \gamma_\lambda (1-\gamma_5)
u_1^\nu (P_1) \{ -F_T g_{\mu \nu }
 +  {1\over 1-\omega^2} (F_L - \omega F_T) v_{1\mu } v_{2\nu } \}
\end{eqnarray}
where $\omega = v_1.v_2 $ is the new variable used instead of $ -q^2 =
(P_1 -P_2)^2 $. The form factors $F_L,F_T$ are combinations of $F_1,F_2$:
\begin{eqnarray}
 F_L & = &\omega F_1 + ( 1 - \omega^2 ) F_2 \nonumber\\
 F_T & = & F_2
\end{eqnarray}
The invariant form factors $F_\Lambda, F_1$ and $F_2 $ are given by the loop
integrals
\begin{eqnarray}
F_\Lambda (\omega )& = &\int d^4k_1d^4k_2[(\not v_2+1)\gamma_5 C]^{+\delta
\sigma }
A^{+\beta \gamma }_{\phantom{=+} \delta \sigma }(v_2,k_1,k_2) \nonumber\\
& ( &\not k_1 - m_1)^ {\beta '}_\beta (\not k_2 -m_2)^{\gamma '}_\gamma
A^{\delta '\sigma '}_{\beta '\gamma '}(v_1,k_1,k_2) [(\not v_1+1)
\gamma_5 C ]_{\delta '\sigma '}\\
 F_1(\omega ) g_{\mu \nu } & + & F_2(\omega ) v_{1\mu } v_{2\nu }= \int d^4k_1
d^4k_2 [(\not v_2+1)\gamma_\mu C]^{+\delta \sigma } A_{\phantom{+}\delta \sigma
}^
{+\beta \gamma } (v_2,k_1,k_2) \nonumber \\
& ( &\not k_1 - m_1)^{\beta'}_\beta (\not k_2 - m_2)_
\gamma ^{\gamma'} A^{\delta'\sigma'}_{\beta'\gamma'}(v_1,k_1,k_2)[(\not v_1+1)
\gamma_\nu C]_{\delta'\sigma'}
\end{eqnarray}
The translational invariance insures that the form factors should depend only
on the variable $\omega $ ( or $ q^2 $).
\subsection{{\it The heavy to light baryon transitions}}
Using the general ansatz (2.3) for light baryon wave functions and the heavy
baryon wave functions we can calculate the following decay matrix elements:\\

i) $\Lambda_Q (\Xi_Q ) \rightarrow \Lambda_q(\Xi_q)$, ${1\over 2}^+
\rightarrow {1\over 2}^+$ baryon transitions
\begin{equation}
< \Lambda_q(P_2) |
J^{V-A}_\lambda | \Lambda_Q(P_1) > = b(\Lambda_q,\Lambda_Q) \bar u_2(P_2)
[ F^1_\Lambda + \not v_1 F^2_\Lambda] \gamma_\lambda (1-\gamma_5)
u_1(P_1).F_\Lambda
\end{equation}
 where
\begin{eqnarray}
\Big (F_\Lambda^1 + \not v_1
F_\Lambda^2 &+& \not v_2 F_\Lambda^3 + \not v_2 \not v_1 F_\Lambda^4)^\alpha
_\rho = \int d^4k_1d^4k_2 [(\not v_2+1)\gamma_5 C]^{+\delta \sigma }
A^{+\alpha \beta \gamma }_{\phantom{+}\rho \delta \sigma }(v_2,k_1,k_2)
\nonumber \\
&(&\not k_1 - m_1)_\beta^ {\beta'} (\not k_2 - m_2)_\gamma ^{\gamma'}
A^{\delta' \sigma'}_{\beta'\gamma'}(v_1,k_1,k_2) [(\not v_1 + 1)\gamma_5
C]_{\delta'\sigma'}
\end{eqnarray}
The terms containing $\not v_2$ and $\not v_2 \not v_1 $ collapse because
of the action on the spinor $ \bar u_2$.

ii) $\Sigma_Q (\Omega_Q) \rightarrow \Sigma_q (\Omega_q) , {1\over 2}^+
\rightarrow {1\over 2}^+$ transitions
\begin{eqnarray}
 <\Sigma_q(P_2) |& J_\lambda^{V-A}& | \Sigma_Q(P_1) >  \nonumber\\
                   & = & b(\Sigma_Q,\Sigma_q) \bar u_2(P_2) \gamma_5
(\gamma^\mu + v_2^\mu )L_{\mu \nu } \gamma_\lambda
(1-\gamma_5)(\gamma^\nu + v^\nu_1) \gamma_5 u_1(P_1)\nonumber\\
& = & 2
b(\Sigma_Q,\Sigma_q)\bar u_2(P_2)\gamma_5[\bar F_1\gamma_\nu + \bar F_2
v_{2\nu } + \bar F_3 \gamma _\nu \not v_1 \nonumber\\
& + & \bar F_4 v_{2\nu } \not v_1]
\gamma_\lambda (1 - \gamma_5)(\gamma^\nu + v_1^\nu )\gamma_5 u_1(P_1)
\end{eqnarray}
where
\begin{eqnarray}
 (L_{\mu \nu })^\alpha _\rho & = & \int d^4k_1d^4k_2
[(\not v_2 + 1 )\gamma _\mu C]^{+ \delta \sigma } A^{+\alpha \beta \gamma
}_{\phantom{+}\rho \delta \sigma } (v_2,k_1,k_2) \nonumber\\
&.&(\not k_1 - m_1)_\beta^{\beta'}. (\not k_2 - m_2)_\gamma^{\gamma '}
A^{\delta' \sigma'}_{\beta'\gamma'}(v_1,k_1,k_2) [(\not v_1 +1)
\gamma_\nu C]_{\delta' \sigma'}
\end{eqnarray}

iii) $\Sigma_Q (\Omega_Q ) \rightarrow \Sigma_q^*(\Omega_q^*) , {1\over 2}^+
\rightarrow {3\over 2}^+$ transitions
\begin{eqnarray}
<\Sigma_q^*(P_2)|& J_\lambda^{V-A}&| \Sigma_Q(P_1) > = b(\Sigma_Q,\Sigma^*_q)
\bar u^\mu_2 L_{\mu \nu }\gamma_\lambda ( 1- \gamma_5) (\gamma^\nu + v^\nu_1)
\gamma_5 u_1 (P_1)\nonumber\\
& = & b(\Sigma_q,\Sigma_Q^*)\bar u_2^\mu (P_2)\Big[-G^*_1 g_{\mu \nu } + G^*_2
v_{1\mu }v_{2\nu } + G_3^* g_{\mu \nu }\not v_1 + G_4^* v_{1\mu }v_{2\nu }
\not v_1 \nonumber\\
& + & G^*_5 v_{1\mu }\gamma_\nu + G^*_6 v_{1\mu }\gamma _\nu \not v_1\Big]
\gamma_\lambda (1-\gamma_5)(\gamma^\nu + v_1^\nu )\gamma_5 u_1(P_1)
\end{eqnarray}

iv) $\Sigma^*_Q(\Omega^*_Q)\rightarrow \Sigma^*_Q(\Omega^*_Q)$-type
${3\over 2}^+\rightarrow {3\over 2}^+$ transition
\begin{eqnarray}
< \Sigma_Q^*(P_2) | &J^{V-A}_\lambda & | \Sigma_Q^*(P_1) >=
b(\Sigma^*_1,\Sigma_2^*) \not u_2^\mu (P_2) L_{\mu \nu }\gamma_\lambda
(1- \gamma_5)
u^\nu_1(P_1) \nonumber\\
& = & 4 b(\Sigma_Q^*,\Sigma_Q^*) \not u_2^\mu (P_2)\Big[ ( -G_1^* g_{\mu \nu }
 + G^*_2 v_{1\mu } v_{2\nu })\gamma_\lambda (1-\gamma_5) \nonumber\\
& - & (G^*_3 g_{\mu \nu } +
G_4^* v_{1\mu }v_{2\nu })\gamma _\lambda ( 1 + \gamma_5 )
 + 2(G^*_3 g_{\mu \nu }
+ G_4^* v_{1\mu}v_{2\nu })v_{1\lambda }(1-\gamma_5)\nonumber\\
& + & 2G_5^* v_{1\mu }g_{\nu \lambda }(1- \gamma_5) + 2 G_6^* v_{1\mu } g_{\nu
\lambda } ( 1 + \gamma _5)\Big] u_1^\nu (P_1)
\end{eqnarray}

\section{The covariant Bethe-Salpeter formalism for the quark-diquark systems}
The presence of diquark gives more restrictions to the ansatz given above. Let
us define the concept of diquark in our framework. In order to have a sense,
the
diquark must have a well-defined spin. The spin-spin interaction must be
negligible beside the static color  potential between the diquark and the
quark
to preserve spin as a good quantum number. Mathematically speaking, the
diquark-quark picture is an approximation, where the spin part of the baryon
wave
function can be approximated as follows:
\begin{equation}
 A^{\rho \delta \sigma }_{\alpha \beta \gamma } \approx A^{\rho \delta }_{
\alpha \beta }(v,k_2)D^\sigma _\gamma (v,k_1)
\end{equation}
Hereafter we shall use the variable $v$ instead of $P$. Diagramatically, the
quark-diquark Bethe-Salpeter wave functions are represented in Fig.4\\
\vskip 3truecm
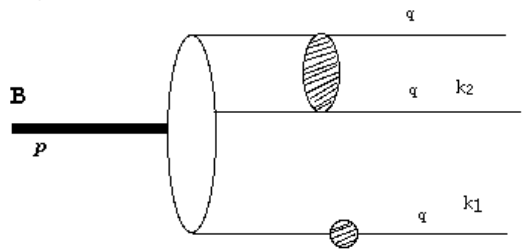
\centerline {{\footnotesize Fig.4}}
Let us notice that the $k_1$-dependence of the diquark propagator and of the
truncated diquark-quark B-S vertex function has been absorbed into $D(
v,k_1)$. So, the light baryon Bethe-Salpeter wave functions are written in the
form:\\

i) for the $\Lambda (\xi )$-type baryons ($J^P = {1\over 2}^+$)
\begin{equation}
\Lambda ^l_{\alpha \beta \gamma }= u_\rho (v)[(\not v +1)\gamma _5 C]_{\delta
\sigma }A^{[\rho \delta ]}_{\alpha \beta }(v, k_2)D^\sigma _\gamma (v, k_1)
\end{equation}

ii) for the $\Sigma ( \Omega )$-type baryons ($J^P = {1\over 2}^+$):
\begin{equation}
 \Sigma^l_{\alpha \beta \gamma } = - [(\gamma ^\mu + v^\mu)\gamma _5 u]_\rho
[(\not v +1)\gamma _\mu C]_{\delta \sigma } A^{(\rho \delta )}_{\alpha \beta }
(v, k_2) D^\sigma _\gamma (v, k_1)
\end{equation}

iii) for the $\Sigma ^*(\Omega ^*)$-type baryons ($J^P = {3\over 2}^+$)
\begin{equation}
 \Sigma ^{*l} = u^\mu_\rho (v) [(\not v +1)\gamma _\mu C]_{\delta \sigma }.
A^{(\rho \delta )}_{\alpha \beta }(v,k_2). D^\sigma _\gamma (v, k_1)
\end{equation}
Specially, if one of quarks becomes heavy, the limit of the heavy quark mass
going to infinity will give further simplification to the above ansatz.
Let us consider two different cases
\subsection {{\it The heavy diquark picture of 1HQ-baryons:}}
This case is achieved by applying both the approximation (3.1) and (2.11).
So, we have
\begin{equation}
 A^{\rho \delta \sigma }_{\alpha \beta \gamma } \approx \delta^\rho_\alpha
A^\delta_\beta (v, k_2) D^\sigma _\gamma(v, k_1)
\end{equation}

Diagramatically, we represent the Bethe-Salpeter amplitude in Fig.5\\

\vskip 3truecm
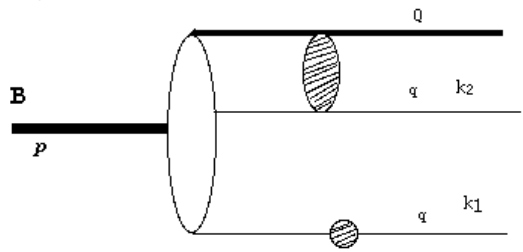
\centerline {{\footnotesize Fig.5}}
We can also use the spin-parity projectors given in (2.4-6) to project out the
baryon B-S wave functions in a similar way as in (3.2-5).
\subsection{{ \it The light diquark picture of 1HQ-baryons:}}
Assuming that two light quarks of a 1HQ-baryon form a light diquark, if we
split
the spin part of the wave function accordingly, the B-S amplitude of the light
diquark-heavy quark system is also represented by the diagram in Fig.3. It
means
that the existence of light diquark in 1HQ-baryons does not lead to any
simplification.

\section { The heavy to heavy current induced baryon transitions in the heavy
diquark model}
Diagramatically, these transitions are represented by the Feynman graph in
Fig.6\\
\vskip 3 truecm
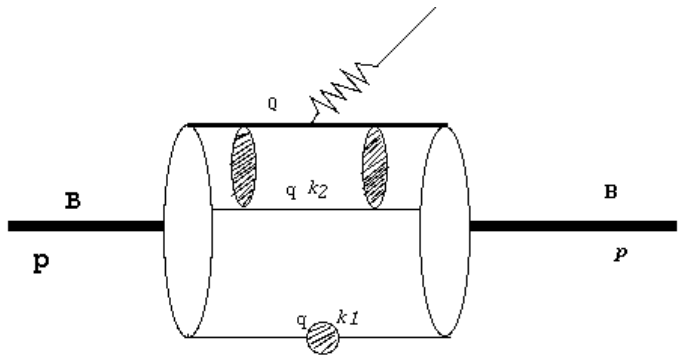
\centerline {{\footnotesize Fig.6 }}
Using the wave functions projected out by the spin-parity projectors (2.4-6)
and the approximation (3.5) let us compute the decay matrix elements for the
heavy to heavy baryon transitions.

The Bethe-Salpeter wave functions for baryons in the heavy diquark model are
follows
\begin{eqnarray}
\Lambda _{Q \alpha \beta \gamma }& = & u_{\alpha }(v) [(\not v + 1)
\gamma _5 C]_{\delta \sigma } A_{\beta }^{\delta }(v,k_2)D_{\gamma }^{\sigma }
(v,k_1)\\
 \Sigma _{Q\alpha \beta \gamma } & = &- [(\gamma ^\mu + v^\mu )
\gamma _5 u]_\alpha [(\not v +1) \gamma _\mu C]_{\delta \sigma }
A^\delta _\beta (v,k_2) D^{\sigma }_{\gamma } (v,k_1) \\
 \Sigma^*_{Q\alpha \beta \gamma } & = & u^\mu _\alpha (v) [( \not v + 1 )
\gamma _\mu C ]_{\delta \sigma } A^\delta_\beta (v,k_2)
D^\sigma_\gamma (v,k_1)
\end{eqnarray}
Let us define the tensor:
\begin{eqnarray}
 A^{\delta'}_\delta (v_1,v_2)& = & \int d^4k_2 A^{+\beta }_{\phantom{+}
\delta }(v_2,k_2) (\not k_2 - m_2 )_\beta ^{\beta'}
A_{\beta'} ^{\delta'}(v_1,k_2) \nonumber \\
& = &( A_1(\omega ) + A_2(\omega ) \not v_1 + A_3(\omega ) \not v_2 \not v_1
+ A_4 (\omega )\not v_2 )^{\delta'}_\delta \\
 D^{\sigma'}_\sigma & = &\int d^4k_1 D^{+\gamma }_{\phantom{+}\sigma }
(\not v_1 - m_1 )_\gamma ^{\gamma'} D_{\gamma'}^{\sigma'} \nonumber\\
& = & ( D_1(\omega ) + D_2(\omega ) \not v_1 + D_3(\omega ) \not v_2 \not v_1
+ D_4(\omega )\not v_2)^{\sigma'}_\sigma
\end{eqnarray}
The matrix elements are computed as follows:\\

i) $\Lambda_Q (\Xi_Q ) \rightarrow \Lambda_Q(\Xi_Q)$, ${1\over 2}^+ \rightarrow
{1\over 2}^+$ baryon transitions
\begin{equation}
< \Lambda_1(P_2) | J^{V-A}_\lambda | \Lambda_(P_1) > =
b(\Lambda_1,\Lambda_2) \gamma_\lambda (1-\gamma_5)
u_1(P_1). \tilde F_\Lambda
\end{equation}
where
\begin{eqnarray}
 \tilde F_\Lambda (\omega ) & = & [(\not v_2 + 1 )\gamma_5 C]^{+\delta
\sigma }[(\not v_1 + 1)\gamma_5C]_{\delta'\sigma'} A^{\delta'}_{\delta }
D^{\sigma'}_\sigma \nonumber \\
& = & Tr([(\not v_2+1)\gamma_5 C]^+[(\not v_1 +1)\gamma_5 C]) A_1(\omega ) D_1
(\omega ) \nonumber \\
& = & 8 A_1(\omega )D_1(\omega )
\end{eqnarray}

ii) $\Sigma_Q (\Omega_Q) \rightarrow \Sigma_Q (\Omega_Q) , {1\over 2}^+
\rightarrow {1\over 2}^+$ transitions
\begin{eqnarray}
<\Sigma_Q(P_2) |& J_\lambda^{V-A}& | \Sigma_Q(P_1) > \nonumber\\
    & = & b(\Sigma_Q,\Sigma_Q) \bar u_2(P_2) \gamma_5 (\gamma^\mu + v_2^\mu)
\gamma_\lambda (1-\gamma_5)(\gamma^\nu + v^\nu_1) \gamma_5 u_1(P_1)
\tilde L_{\mu \nu }
\end{eqnarray}
where
\begin{eqnarray}
 \tilde L_{\mu \nu }& = & [(\not v_2+1)\gamma _\mu C]^{\delta \sigma }
[(\not v_1+1) \gamma _\nu C]_{\delta'\sigma'} A_\delta ^{\delta'}
D_\sigma^{\sigma'}\nonumber\\
& = & Tr([(\not v_2 +1)\gamma_\mu c]^+[(\not v_1)\gamma_\nu C]) A_1(\omega )
D_1(\omega )\nonumber\\
& = & 8 (g_{\mu \nu } -v_{1\mu } v_{2\nu }) A_1(\omega ) D_1(\omega )
\end{eqnarray}

iii) $\Sigma_Q (\Omega_Q ) \rightarrow \Sigma_Q^*(\Omega_Q^*) , {1\over 2}^+
\rightarrow {3\over 2}^+$ transitions
\begin{equation}
<\Sigma_Q^*(P_2) | J_\lambda^{V-A} | \Sigma_Q(P_1) > = b(\Sigma_Q,\Sigma^*_Q)
\bar u^\mu_2 \gamma_\lambda ( 1- \gamma_5) (\gamma^\nu + v^\nu_1)
\gamma_5 u_1 (P_1) \tilde L_{\mu \nu }
\end{equation}

iv) $\Sigma^*_Q(\Omega^*_Q)\rightarrow \Sigma^*_Q(\Omega^*_Q)$-type ${3\over
2}^+
\rightarrow {3\over 2}^+$ transition
\begin{equation}
< \Sigma_Q^*(P_2) | J^{V-A}_\lambda | \Sigma_Q^*(P_1) >=
b(\Sigma^*_1,\Sigma_2^*) \bar u_2^\mu (P_2) \gamma_\lambda (1- \gamma_5)
u^\nu_1(P_1) \tilde L_{\mu \nu }
\end{equation}
So if the heavy diquark picture is a good approximation, all the heavy to heavy
baryon transitions are characterized by only one form factor $ \tilde F(\omega
)
= 8 A_1(\omega ) D_1 (\omega ) $. Comparing with the results given in Sec.II,
we come to the conclusion that the presence of heavy diquark reduces
the number of the weak form factor from 3 to 2. The existence of the heavy
quark is checked by the relations
\begin{eqnarray}
  F_1(\omega ) & = & \omega \tilde F(\omega ) \nonumber\\
  F_2(\omega ) & = & F_\Lambda (\omega ) = \tilde F(\omega )
\end{eqnarray}
or
\begin{equation}
 \omega F_2(\omega ) = F_1(\omega ) =\omega F_\Lambda (\omega )
\end{equation}
Checking the equalities (4.14) in the heavy to heavy baryon transitions we can
be able to verify or exclude the heavy diquark picture. The relations between
weak form factors in this section have been obtained in the work\cite{HU2}
based on the assumption of "independent light quarks". According to our
interpretations given previously, it is the case of the heavy diquark.
\section{The heavy to light transitions in the heavy diquark model of
1HQ-baryons}
There are two cases represented by the Feynman graphs in Fig.7 and Fig.8.
Let us consider these cases separately
\subsection{ {\it The case of stable diquark}}
The Feynman graph in Fig.7 corresponds to this case, where the decaying diquark
remains stable. In this case the correlation inside the diquark remains strong
to keep these quarks together. The produced quark cannot be lighter than the
spectator quark according to the idea of heavy diquark.

\vskip 3cm
\centerline{{footnotesize Fig.7}}
Using the diquark-quark-baryon B-S wave functions of heavy baryons and light
baryons given in Sec.III. Let us consider the $k_2$-dependent part of the
light baryon wave functions
\begin{equation}
   A^{\rho \delta }_{\alpha \beta }(v,k_2) = A^{[\rho \delta ]}_{\alpha
\beta }(v,k_2) + A^{(\rho \delta )}_{\alpha \beta }
\end{equation}
The symmetric part $ A^{(\rho \delta )}_{\alpha \beta }$ can be written as
follows:
\begin{equation}
A^{(\rho \delta )}_{\alpha \beta } = A^1_{\alpha \beta } (C^{-1})^{\rho \delta
}
+ A^2_{\alpha \beta } (C^{-1}\gamma_5)^{\rho \delta } + A^{3\mu }_{\alpha
\beta }
(C^{-1}\gamma _\mu \gamma_5)^{\rho \delta }
\end{equation}
The antisymmetric part is
\begin{equation}
 A^{[\rho \delta ]}_{\alpha \beta } = \tilde A^{1\mu }_{\alpha
\beta }(C^{-1}\gamma _\mu)^{\rho \delta } + \tilde A^{2\mu \nu}_{\alpha \beta }
(c^{-1}
\sigma _{\mu \nu })^{\rho \delta }
\end{equation}
Inserting the expressions (5.2) and (5.3) into the wave function of light
baryons, we can compute the following decay matrix elements\\

i) $\Lambda_Q (\Xi_Q ) \rightarrow \Lambda_q(\Xi_q), {1\over 2}^+ \rightarrow
{1\over 2}^+$ baryon transitions
\begin{eqnarray}
< \Lambda_q(P_2) |&J^{V-A}_\lambda &| \Lambda_Q(P_1) > = b(\Lambda_q,\Lambda_Q)
\bar u_2^\rho (P_2) [\gamma_\lambda (1-\gamma_5) u_1(P_1)]_\alpha \hat
F_\rho ^\alpha \nonumber\\
& = & \bar u_2(P_2) (F^\Lambda_1(\omega ) + \not v_1 F^\Lambda_2(\omega ))
\gamma_\lambda (1-\gamma_5)u(v_1)
\end{eqnarray}
where
\begin{equation}
 \hat F^\alpha _\rho =[(\not v_2+1)\gamma_5 C]^{+\delta \sigma }[(\not v_1 +1)
\gamma_5 C]_{\delta'\sigma }\hat A^{\alpha \delta'}_{[\rho \delta ] }
D_1(\omega )
\end{equation}
and
\begin{equation}
 \hat A ^{\alpha \delta'}_{[\rho \delta ]} = \int d^4k_2 A^{+\alpha \beta }_{
\phantom{+}\rho \delta }(v_2,k_2)(\not k_2 - m_2)^{\beta'}_\beta
A_{\beta'}^{\delta'}(v_1,v_1)
\end{equation}
There is no collapsing of the form factors. However, the loop integral is
greatly simplified.

ii) $\Sigma_Q (\Omega_Q) \rightarrow \Sigma_q (\Omega_q) , {1\over 2}^+
\rightarrow {1\over 2}^+$ transitions
\begin{eqnarray}
<\Sigma_q(P_2) |&J_\lambda^{V-A}&| \Sigma_Q(P_1) >=b(\Sigma_Q,\Sigma_q)
\big [\bar u_2(P_2) \gamma_5 (\gamma^\mu + v_2^\mu)\big]^\rho \big[\gamma_
\lambda (1-\gamma_5)(\gamma^\nu   \nonumber\\
&+& v^\nu_1) \gamma_5 u_1(P_1)\big]_\alpha
[(\not v_2 +1)\gamma_\mu C]^{+\delta \sigma }[(\not v_1 +1)\gamma_\nu C]_{
\delta'\sigma }\tilde A^{\alpha \delta'}_{(\rho \delta )} D_1(\omega )
\nonumber \\
& = & \bar u_2(P_2)(\not v_1+1)\gamma_\nu \gamma_5(\tilde G_1(\omega ) +
\tilde G_2(\omega ) \not v_2 )\gamma _\lambda (1-\gamma_5)(\gamma ^\nu +
v_1^\nu )\gamma_5 u_1(P(1)
\end{eqnarray}
where
\begin{equation}
\tilde A^{\alpha \delta'}_{(\rho \delta )}=\int d^4k_2 A^{+\alpha \beta }_
{\phantom{+} (\rho \delta )}(v_2,k_2)(\not k_2 - m_2)^{\beta'}_\beta
A_{\beta'}^{\delta'}(v_1,k_2)
\end{equation}
Comparing with the formulas given in Sec.II we obtain the relations
\begin{eqnarray}
 F_1(\omega )& = & -\tilde G_1(\omega ) -\tilde G_2(\omega ) - 2 \omega \\
 F_2(\omega )& = & - 2\tilde G_2(\omega )  \\
 F_3(\omega )& = & \tilde G_2(\omega ) - \tilde G_1(\omega ) \\
 F_4(\omega )& = & 2\tilde G_2(\omega )
\end{eqnarray}
The relations to be verified are
\begin{eqnarray}
 F_2(\omega )& = & -F_4(\omega ) \\
 F_1(\omega )& = & F_3(\omega ) -F_4(\omega ) - 2\omega
\end{eqnarray}
Four form factors collapse to two.

iii) $\Sigma_Q (\Omega_Q ) \rightarrow \Sigma_q^*(\Omega_q^*) , {1\over 2}^+
\rightarrow {3\over 2}^+$ transitions
\begin{eqnarray}
<\Sigma_q^*(P_2) |& J_\lambda^{V-A}& | \Sigma_Q(P_1) > = b(\Sigma_Q,
\Sigma^*_q)
(\bar u^{\mu})^\rho_2\Big [\gamma_\lambda ( 1- \gamma_5) (\gamma^\nu + v^\nu_1)
\nonumber\\
& &\gamma_5 u_1 (P_1)\Big]^\alpha
\big[(\not v_2 +1)\gamma_\mu C\big]^{+\delta \sigma }\big[(\not v_1+1)
\gamma_\mu C\big]_{\delta' \sigma } \tilde A^{\alpha \beta }_{(\rho \delta )}
(v_1,v_2)D_1(\omega )\\
& = & b(\Sigma_q,\Sigma_Q^*)\bar u_2^\mu (P_2)\gamma_\nu\big[ v_{1\mu}
H_1(\omega ) +
H_2(\omega ) v_{1\mu} \not v_2\big] \gamma_\lambda(1-\gamma_5)(\gamma^\nu
\nonumber \\
& + & v_1^\nu)\gamma_5 u_1(P_1)
\end{eqnarray}
Comparing with the formulas in Sec.II. we obtain the relations
\begin{eqnarray}
G^*_1(\omega ) & = & 0 \nonumber \\
G_2^*(\omega ) & = & 2H_2(\omega )(2\omega -1)\nonumber \\
G_3^*(\omega ) & = & 0 \nonumber \\
G_4^*(\omega ) & = & -2 H_2(\omega ) \nonumber \\
G_5^*(\omega ) & = & (1-2\omega )H_2(\omega ) -H_1(\omega ) \nonumber \\
G_6^*(\omega ) & = & H_1(\omega ) + H_2(\omega )
\end{eqnarray}
The number of the form factors has also reduced to 2.\\

iv) $\Sigma^*_Q(\Omega^*_Q)\rightarrow \Sigma^*_Q(\Omega^*_Q)$-type
${3\over 2}^+ \rightarrow {3\over 2}^+$ transition
\begin{eqnarray}
< \Sigma_Q^*(P_2) |& J^{V-A}_\lambda & | \Sigma_Q^*(P_1) >=
b(\Sigma^*_1,\Sigma_2^*) (\bar u_2^\mu )^\rho (P_2) [\gamma_\lambda (1-
\gamma_5) u^\nu_1(P_1)]_\alpha \nonumber \\
& & [(\not v_2 +1)\gamma _\mu C]^{+\delta \sigma}[(\not v_1+1)\gamma_\nu C]_
{\delta'\sigma } \tilde A^{\alpha \delta'}_{(\rho \delta ) } D_1(\omega
)\nonumber \\
& = & \bar u_2^\mu (v_2) \gamma^\nu (\not v_1 -1)v_{1\mu }(H_1(\omega ) +
H_2(\omega ) \not v_2)\gamma_\lambda(1-\gamma_5) u_1(P_1)
\end{eqnarray}
We have also two form factors $ H_1$ and $H_2$ in this case
\subsection{{\it The case of new light diquark forming in the produced baryon}}
This case is represented by the Feynmann graph in Fig.8.

\vskip 3truecm
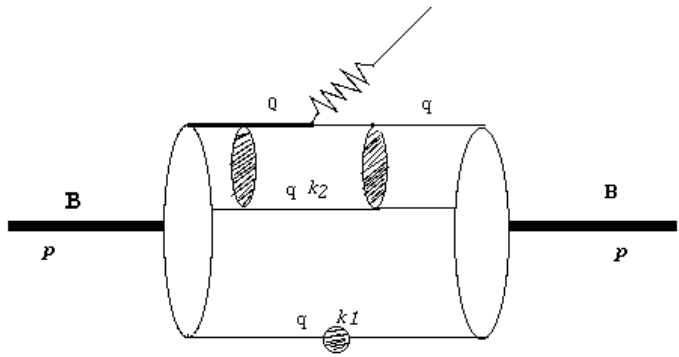
\centerline {{\footnotesize Fig.8 }}
In this case the new quark becomes lighter than the spectator one, so the
diquark becomes unstable and one of its quark leaves the boundstate to join
the other quark forming a new light diquark. Inserting the wave functions
of light baryons and heavy baryons into the formula (2.12) we see that as the
loop integrals are not separated no collapsing of the form factors occurs as
consequence of the diquark picture.
\section{Discussion}
Let us summarize the results before going further. The suggestion that the
heavy diquark may exist
inside the 1HQ-baryons leads to some constraints between the weak form factors
in some cases. It is remarkable that the concept of diquark used in this paper
is after Lichtenberg[LICH]. That means, we have not postulated more than
to suppose
that the correlation between two quarks is much stronger than any other. No
local elementary diquark field has been assumed. The analysis here has been
based on
the fully antisymmetrized wave function of baryons.
In the heavy to heavy baryon transitions the number of form
factors reduces from 3 to 1 as the consequence of heavy diquark hypothesis.
In the heavy to light baryon transitions if the produced quark can still keep
its partner the number of the form factors reduces greatly. The stability of
the decaying diquark could have different reasons. In our opinion if
the produced quark is still heavier than the quark outside the diquark it will
be the case. Anyway, the constraints on the weak form factors of these
transitions
would testify which diquark exists in the produced baryons.

 If in the heavy baryon a point-like heavy diquark does exist we should be able
to indicate its existence by the structure of form factors. Assuming that
such a point-like structure does
exit inside baryons, we can treat the diquark as an elementary local field.
As the diquark is a boson and very heavy, in the limit of the heavy quark mass
going to infinity we can construct an effective theory for heavy
diquark following Georgi, Wise and Carone\cite{GEWICA}. Now the heavy boson of
Georgi, Wise and
Carone has a new interpretation as the heavy diquark. It is interesting to note
that in such an effective theory in the heavy quark mass limit, there is a
supersymmetry between the heavy diquark and the heavy antiquark. If this
supersymmetry survives the hadronic level, we can use the Wigner-Eckart
theorem to derive the same formula as the one of Georgi and Carone\cite{GEWICA}
for the matrix element:
\begin{equation}
 < \Lambda_Q(P_2) | J^{V}_\lambda | \Lambda_Q(P_1) > = b(\Lambda_q,\Lambda_Q)
\bar u_2(v_2)
u_1(v_1)\xi(\omega )(v_{1\mu }+v_{2\mu } )
\end{equation}
The $\xi (\omega )$ form factors is sometimes called Isgur-Wise function. This
result is quite different from the formula (2.13). By the help of the Gorkov
identity we can transform the formula (6.1) into another form with the electric
-type term containing $\gamma_\lambda $ and a magnetic term containing $\sigma_
{\lambda \mu }(v_2^\mu -v_1^\mu )$. Comparing with the equation (2.13), we see
that an elementary heavy diquark structure leads to a magnetic form factor,
while the usual HQET does not.
 The explanation is simple: the magnetic moment
of the baryon with an elementary diquark comes from the spin of the light
quark.
With light diquark the magnetic moment of the baryon comes from the spin of the
heavy quark, which contributes nothing to the weak form factors due to the
heavy
quark symmetry.

It is also interesting to note that the heavy diquark picture give the same
number of form factors for mesons and baryons. This would mean that the
supersymmetry can occur. Originally, supersymmetry of hadrons was proposed
by Miyazawa\cite{MIYA} to derive some similarities between baryons and mesons
as
early as in 1968. Recently Catto and G\"ursey\cite{CAGU}found QCD basics for
this
symmetry and explained the parallelism between mesonic and baryonic Regge
trajectories. As the supersymmetry is broken badly for light hadrons,
Lichtenberg\cite{LISUSY} argued that the supersymmetry will become good for
heavy hadrons. There are two types of supersymmetry depending on the existence
of
diquark. The supersymmetry of Lichtenberg is associated with the light diquark.
The second supersymmetry $SU(1/6)$ is associated with the heavy quark. So far,
the supersymmetry in the world of 1HQ-baryons
occurs on two basic assumptions : the existence of diquark and the heavy
quark symmetry. It seems too strict to suppose so much. But the nature is
always
simpler than we thought. The question is whether the heavy diquark picture is
good and on which energy scale. It is very likely that the heavy diquark does
exist on the same energy
scale as the one of the heavy quark symmetry or even lower than that. If this
is
true we can expect that the supersymmetry occurs at the same time as the heavy
quark symmetry.

Presently, we don't have much informations about weak decays of heavy baryons.
Hopefully, more data of heavy baryon's weak decays will be available in the
near future to
clarify our questions: Which diquark picture ( heavy or light) is the better
approximation to 1HQ baryons? Can diquark be approximated as a point-like
object? Is there any supersymmetry between 1HQ hadrons? Basing on our results
a future analysis on the weak decay data of
baryons will determine whether these postulations can make any sense.

\section{Acknowledgements}

  Many thanks are due to Professor F.Hussain for his interest in this work.
Many criticisms by him has been taken into the paper and the main idea was
also initiated by the work\cite{HU1} of his group.

  We should also like to acknowledge the International Centre for Theoretical
Physics for the hospitality.

  Thanks are due to Professor K.C.Wali for the hospitality at Syracuse
University, where this work has been completed.

   Professor S.Fredriksson provided us a wide list of references on diquark,
 which undoubtfully has sharpenned our understanding of diquark.

\newpage
\figure{ Fig.1 The quark-baryon Bethe-Salpeter amplitude. }

\figure{ Fig.2 The 1HQ baryon-quark Bethe-Salpeter amplitude.}

\figure{ Fig.3 The weak current induced baryon transitions.}

\figure{ Fig.4 The ansatz for the light baryon Bethe-Salpeter amplitudes}

\figure{ Fig.5 The heavy diquark-quark-baryon Bethe-Salpeter amplitude}

\figure{ Fig.6 The heavy to heavy current induced baryon transition in the
heavy diquark model}

\figure{ Fig.7 The heavy to light baryon transition in the heavy diquark model.
The diquark remains stable after the weak decay of the heavy quark}

\figure{ Fig.8 The heavy to light baryon transition in the heavy diquark model.
A new diquark appears inside the light baryon }

\end{document}